\documentclass[reprint,amsmath,amssymb,aip,jcp]{revtex4-1}

\usepackage{graphicx}
\usepackage{dcolumn}
\usepackage{bm}

\usepackage{amsmath}
\usepackage{amsbsy}
\usepackage{color}

\newcommand{\be}{\begin{equation}}
\newcommand{\ee}{\end{equation}}
\newcommand{\ba}{\begin{eqnarray}}
\newcommand{\ea}{\end{eqnarray}}
\newcommand{\nn}{\nonumber\\}
\newcommand{\ocite}{\onlinecite}

\newcommand{\widthfigure}{.7}

\newcommand{\xx}{5}
\newcommand{\yy}{5}

\newcommand{\xmax}{40}
\newcommand{\ymax}{22}

\newcommand{\DA}{\begin{picture}(\xmax,\ymax)(-\xx,\yy)
\setlength{\unitlength}{.1mm}
\put(0,60){\circle {18}}
\put(60,60){\circle {18}}
\put(0,0){\circle {18}}
\put(60,0){\circle*{18}}
\put(9,60){\line(1,0){42}}
\put(9,0){\line(1,0){42}}
\put(0,9){\line(0,1){42}}
\put(60,9){\line(0,1){42}}
\end{picture}}

\newcommand{\DB}{\begin{picture}(\xmax,\ymax)(-\xx,\yy)
\setlength{\unitlength}{.1mm}
\put(0,60){\circle {18}}
\put(60,60){\circle {18}}
\put(0,0){\circle {18}}
\put(60,0){\circle*{18}}
\put(9,60){\line(1,0){42}}
\put(9,0){\line(1,0){42}}
\put(0,9){\line(0,1){42}}
\put(60,9){\line(0,1){42}}
\put(7,7){\line(1,1) {46.5}}
\end{picture}}

\newcommand{\DC}{\begin{picture}(\xmax,\ymax)(-\xx,\yy)
\setlength{\unitlength}{.1mm}
\put(0,60){\circle {18}}
\put(60,60){\circle {18}}
\put(0,0){\circle {18}}
\put(60,0){\circle*{18}}
\put(9,60){\line(1,0){42}}
\put(9,0){\line(1,0){42}}
\put(0,9){\line(0,1){42}}
\put(60,9){\line(0,1){42}}
\put(7,53){\line(1,-1) {46.5}}
\end{picture}}

\newcommand{\DD}{\begin{picture}(\xmax,\ymax)(-\xx,\yy)
\setlength{\unitlength}{.1mm}
\put(0,60){\circle {18}}
\put(60,60){\circle {18}}
\put(0,0){\circle {18}}
\put(60,0){\circle*{18}}
\put(9,60){\line(1,0){42}}
\put(9,0){\line(1,0){42}}
\put(0,9){\line(0,1){42}}
\put(60,9){\line(0,1){42}}
\put(7,7){\line(1,1) {46.5}}
\put(7,53){\line(1,-1) {46.5}}
\end{picture}}

\newcommand{\DE}{\begin{picture}(\xmax,\ymax)(-\xx,\yy)
\setlength{\unitlength}{.1mm}
\put(0,60){\circle*{18}}
\put(60,60){\circle {18}}
\put(0,0){\circle*{18}}
\put(60,0){\circle{18}}
\put(9,60){\line(1,0){42}}
\put(9,0){\line(1,0){42}}
\put(0,9){\line(0,1){42}}
\put(60,9){\line(0,1){42}}
\end{picture}}

\newcommand{\DF}{\begin{picture}(\xmax,\ymax)(-\xx,\yy)
\setlength{\unitlength}{.1mm}
\put(0,60){\circle*{18}}
\put(60,60){\circle {18}}
\put(0,0){\circle{18}}
\put(60,0){\circle*{18}}
\put(9,60){\line(1,0){42}}
\put(9,0){\line(1,0){42}}
\put(0,9){\line(0,1){42}}
\put(60,9){\line(0,1){42}}
\end{picture}}

\newcommand{\DG}{\begin{picture}(\xmax,\ymax)(-\xx,\yy)
\setlength{\unitlength}{.1mm}
\put(0,60){\circle*{18}}
\put(60,60){\circle{18}}
\put(0,0){\circle*{18}}
\put(60,0){\circle{18}}
\put(9,60){\line(1,0){42}}
\put(9,0){\line(1,0){42}}
\put(0,9){\line(0,1){42}}
\put(60,9){\line(0,1){42}}
\put(7,7){\line(1,1) {46.5}}
\end{picture}}

\newcommand{\DI}{\begin{picture}(\xmax,\ymax)(-\xx,\yy)
\setlength{\unitlength}{.1mm}
\put(0,60){\circle*{18}}
\put(60,60){\circle{18}}
\put(0,0){\circle{18}}
\put(60,0){\circle*{18}}
\put(9,60){\line(1,0){42}}
\put(9,0){\line(1,0){42}}
\put(0,9){\line(0,1){42}}
\put(60,9){\line(0,1){42}}
\put(7,7){\line(1,1) {46.5}}
\end{picture}}

\newcommand{\DK}{\begin{picture}(\xmax,\ymax)(-\xx,\yy)
\setlength{\unitlength}{.1mm}
\put(0,60){\circle*{18}}
\put(60,60){\circle{18}}
\put(0,0){\circle{18}}
\put(60,0){\circle*{18}}
\put(9,60){\line(1,0){42}}
\put(9,0){\line(1,0){42}}
\put(0,9){\line(0,1){42}}
\put(60,9){\line(0,1){42}}
\put(7,53){\line(1,-1) {46.5}}
\end{picture}}

\newcommand{\DL}{\begin{picture}(\xmax,\ymax)(-\xx,\yy)
\setlength{\unitlength}{.1mm}
\put(0,60){\circle*{18}}
\put(60,60){\circle{18}}
\put(0,0){\circle*{18}}
\put(60,0){\circle{18}}
\put(9,60){\line(1,0){42}}
\put(9,0){\line(1,0){42}}
\put(0,9){\line(0,1){42}}
\put(60,9){\line(0,1){42}}
\put(7,7){\line(1,1) {46.5}}
\put(7,53){\line(1,-1) {46.5}}
\end{picture}}

\begin{document}



\title{Fourth virial coefficients of asymmetric nonadditive hard-disk mixtures}


\author{Franz Saija}
\email{saija@me.cnr.it}
\affiliation{CNR-IPCF, Viale F. Stagno d'Alcontres, 37-98158, Messina, Italy}

\author{Andr\'es Santos}

\email{andres@unex.es}
\homepage{http://www.unex.es/eweb/fisteor/andres/}

\author{Santos B. Yuste}
\email{santos@unex.es}
\homepage{http://www.unex.es/eweb/fisteor/santos/}

\affiliation{Departamento de F\'{\i}sica, Universidad de
Extremadura, Badajoz, E-06071, Spain}

\author{Mariano L\'opez de Haro}
\email{malopez@servidor.unam.mx}

\affiliation{Centro
de Investigaci\'on en Energ\'{\i}a, Universidad Nacional Aut\'onoma
de M\'exico (U.N.A.M.), Temixco, Morelos 62580, M{e}xico}
\date{\today}

\begin{abstract}
The fourth virial coefficient of asymmetric nonadditive binary mixtures of hard disks is computed with a standard Monte Carlo method. Wide ranges of size ratio ($0.05\leq q\leq 0.95$) and  nonadditivity ($-0.5\leq \Delta\leq 0.5$) are covered. A comparison is made between the numerical results and those that follow from some theoretical developments. {The possible use of these data in the derivation of new equations of state for these mixtures is illustrated by considering a rescaled virial expansion truncated to fourth order. The numerical results obtained using this {equation of state} are compared with Monte Carlo simulation data in the case of {a size ratio $q=0.7$ and two nonadditivities $\Delta=\pm 0.2$}.}
\end{abstract}

\date{\today}


\maketitle
\section{Introduction}
\label{intro}

The key role that hard-core model systems play in liquid state theory is undeniable. This is mostly due to the well-known fact that in some cases it is possible to derive exact and approximate analytical results for their thermodynamic and structural properties.\cite{M08} Moreover, the structural properties of real dense fluids depend essentially on the short ranged repulsive intermolecular forces, which are adequately accounted for by hard-core models in which molecules have no interactions at separations larger than a given distance and experience infinite repulsion if their separation is less than that distance. While pure one-component hard-core systems lead to a fluid-solid transition, mixtures may display more complex phase behavior. For these latter, one can either assume that they are additive, namely that the closest distance of approach of molecules of two different species is the arithmetic mean of the distances between like pairs, or nonadditive, in which the previous condition does not hold. Additive systems have received most of the attention, but the inclusion of nonadditivity, which may be either positive or negative, attempts to incorporate some features of non-hard forces, such as attractions and soft repulsions, into the description. Amongst other things, nonadditivity serves to account for homo-coordination or hetero-coordination in the compositional order of a mixture and also for fluid-fluid demixing. This makes the nonadditive hard-core models of mixtures both attractive and rather versatile and so it is not surprising that they have been the subject of recent attention in the literature. Some examples concerning nonadditive hard spheres (NAHS) may be found in Refs.\ \ocite{PSCG06,MV07,HS10,HS11}.

As far as  mixtures of nonadditive hard disks (NAHD) are concerned, which are the subject matter of this paper, publications are less numerous than in the case of NAHS. However, interest in these model systems, which dates back at least to the late 1970s, has recently experienced a revival. Applications include {lipid monolayers spread on air-water interfaces,\cite{FZM91}} liquid-liquid demixing in a physisorbed mixture of Argon, Krypton, or Xenon on graphite,\cite{MC94} a model for ganglioside lipid and phospholipid interactions in connection with the binding of cholera-toxin to a lipid membrane,\cite{FK03} the morphology of composite latex particles,\cite{DV05} two-dimensional magnetic colloid mixtures,\cite{HLL06} and the asphaltene flocculation inhibition phenomenon.\cite{BOBZD08}

A  binary mixture of NAHD is characterized by the impenetrable diameters of the two species $\sigma_{11}=\sigma_1$ and $\sigma_{22}=\sigma_2$ and by a crossed diameter $\sigma_{12} = \frac{1}{2}(\sigma_{1} + \sigma_{2})(1 + \Delta)$, where the dimensionless parameter $\Delta$ accounts for deviations of the inter-species interactions from additivity.\cite{SFG98} Like in the NAHS model,  the binary mixture shows a tendency to form hetero-coordinated clusters for negative values of the nonadditivity parameter ($\Delta < 0$). On the other hand, for  positive non-additivity ($\Delta > 0$), the system tends to segregate into two fluid phases, one richer in particles of species 1 and the other richer in particles of species 2, respectively.\cite{SG02} On the {computational} side, Dickinson\cite{D77,D79,D80} reported molecular dynamics simulations of NAHD mixtures in which he computed the compressibility factor and the radial distribution functions for a few size ratios and some nonadditivities. Tenne and Bergmann\cite{TB78} developed a scaled-particle theory (SPT) for NAHD mixtures which was later corrected by Bearman and Mazo\cite{BM88,BM89,BM90} in their study of fluid-fluid phase equilibria for positive nonadditivity. The compressibility factors and part of the coexistence curve arising from the SPT were compared to molecular dynamics simulations of an equimolar symmetric mixture of NAHD by Ehrenberg {\it et al.}\cite{ESH90} Singh and Sinha\cite{SS83} used thermodynamic perturbation theory to compute the Helmholtz free energy per particle, the compressibility factor, and the radial distribution function of binary NAHD mixtures with both positive and negative nonadditivity, while Mishra and Sinha\cite{MS85} derived the excess thermodynamic properties of binary  NAHD mixtures including quantum corrections. Nielaba and coworkers\cite{N96,ISN97,N97,N00} combined the Gibbs ensemble Monte Carlo (GEMC) method and finite-size scaling to study demixing of a symmetric NAHD mixture. Hamad and his collaborators\cite{AEH99,HY00} developed equations of state for NAHD mixtures and performed molecular dynamics simulations for a variety of size ratios and values of the nonadditivity. Saija and Giaquinta\cite{SG02} reported Monte Carlo (MC) results for the thermodynamic and structural properties of a symmetric NAHD mixture for positive nonadditivity and studied phase separation for some positive values of the nonadditivity. Depletion interactions in NAHD mixtures were considered by Casta\~neda-Priego {\it et al.},\cite{CRM03} who also indicated that this model may mimic the qualitative features of effective potentials of hard and soft particles. To cope with large nonadditivities, Buhot\cite{B05} used a cluster algorithm to study phase separation of symmetric binary NAHD mixtures, while Gu\'aqueta\cite{G09} used a combination of MC techniques to determine the location of the critical consolute point of asymmetric NAHD mixtures for a wide range of size ratios and values of the positive nonadditivity. More recently, Mu\~noz-Salazar and Odriozola\cite{MO10} used a semi-grand canonical ensemble Monte Carlo method to obtain the fluid-fluid coexistence curve for a symmetric mixture of NAHD and a single positive nonadditivity.

In 2005 three of us\cite{SHY05} introduced an approximate equation of state for nonadditive hard-core systems in $d$ dimensions and, taking $d=2$, compared the results obtained for the corresponding compressibility factor with simulation data. Later, a unified framework for some of the most important theories (including some generalizations) of the equation of state of $d$-dimensional nonadditive hard-core mixtures was presented.\cite{SHY10} The framework was used for $d=3$ to compare the results of the different approaches with simulation data for the fourth virial coefficients that had recently been derived\cite{PCGS07} and with simulation data for the compressibility factor. It was also used to examine the issue of fluid-fluid demixing.

More recently, another of us\cite{S11} computed the fourth virial coefficient of \emph{symmetric} NAHD mixtures over a wide range of nonadditivity. He also compared the fluid-fluid coexistence curve derived from two equations of state built using the new virial coefficients with some simulation results.

One of the major aims of this paper is to present the results of computations of the fourth virial coefficient of \emph{asymmetric} NAHD mixtures, \emph{i.e.}, mixtures such that the size ratio $q = \sigma_{2} / \sigma_{1}$ is different from unity. We will explore a wide range of values of the nonadditivity parameter $\Delta$ and size ratio $q$. These results complement the ones already published for symmetric mixtures\cite{S11} and will afterwards be used to assess the merits and limitations of some theoretical approaches.

The paper is organized as follows. In section \ref{sec2} we provide the known analytical results for the second and third virial coefficients of a NAHD mixture, as well as the graphical representation of the (partial) composition-independent fourth virial coefficients. The approximate theoretical expressions considered in this paper for the fourth virial coefficients are presented in section \ref{sec3}. This is followed in section \ref{sec4} by the results of the MC evaluation of the fourth virial coefficients for a wide range of size ratios and values of the nonadditivity parameter. A comparison of the theoretical approximations with these data is also presented.
{In section \ref{EOS} the equation of state resulting from a rescaled virial expansion truncated to fourth order, as well as the theoretical approximations mentioned above, are compared with new Monte Carlo simulation data in the case of  two mixtures with negative and positive nonadditivities, respectively.} The paper is closed in section \ref{sec5} with some concluding remarks.

\section{Virial coefficients}
\label{sec2}
The virial expansion can be written as
\begin{equation}
\beta P = \rho + B \rho^2 + C \rho^3 + D \rho^4 + \cdots,
\label{s:bpr}
\end{equation}
$P$ is the pressure, $\beta$ is the inverse temperature
in units of the Boltzmann constant and $\rho = \rho_1 + \rho_2$ is the total number density, $\rho_i$ being the partial number  density of species $i$. In a mixture, at variance with the
one-component case, the virial coefficients $B, C, D, \ldots$ do
also depend on the relative concentration of the two species and on the hard-core diameters. The coefficients $B$ and $C$ are exact and well known (see, for instance, Refs.\ \ocite{AEH99,SHY05}). They are given by
\be
B=B_{11}x_{1}^2+2B_{12}x_{1} x_2+B_{22}x_2^2,
\ee
\be
C=C_{111}x_{1}^3+3C_{112}x_{1}^2x_2+3C_{122}x_{1}x_2^2+C_{222}x_2^3,
\ee
where $x_{1}=\rho_1/\rho$ and $x_2=\rho_2/\rho=1-x_1$ are the mole fractions of species $1$ and $2$, respectively. The other quantities read
\begin{equation}
B_{ij}=\frac{\pi}{2}\sigma_{ij}^2,
\label{n1x}
\end{equation}
\be
C_{111}=\frac{\pi^2}{16}b_3\sigma_{1}^4,
\label{H.9a}
\ee
\be
C_{112}=\frac{\pi^2}{16}b_3\sigma_{1}^4 F\left(\frac{\sigma_{12}}{\sigma_{1}}\right),
\label{H.9b}
\ee
\be
C_{122}=\frac{\pi^2}{16}b_3\sigma_{2}^4 F\left(\frac{\sigma_{12}}{\sigma_{2}}\right),
\label{H.9c}
\ee
\be
C_{222}=\frac{\pi^2}{16}b_3\sigma_{2}^4,
\label{H.9d}
\ee
where $b_3=\frac{16}{3}-\frac{4\sqrt{3}}{\pi}\simeq 3.12802$ and the function $F(x)$ is given by
\be
F(x)= \frac{1}{3}G(x)+\frac{2}{3}x^2H(x)
\label{H.13}
\ee
with
\be
G(x)= \frac{4}{\pi b_3}\left(4x^2\cos^{-1} \frac{1}{2x}-\sqrt{4x^2-1}\right),
\label{H.3}
\ee
\be
H(x)= \frac{4}{\pi b_3}\left[2\pi x^2-2\left(2x^2-1\right)\cos^{-1} \frac{1}{2x}-\sqrt{4x^2-1}\right]
\label{H.4}
\ee
for $x\geq \frac{1}{2}$ and
\be
G(x)=0,\quad H(x)=\frac{8}{b_3}x^2
\label{H.4b}
\ee
for $0\leq x\leq \frac{1}{2}$.

In turn, the fourth-order virial coefficient reads
\ba
D &=& D_{1111}x_1^4 + 4D_{1112}x_1^3x_2 + 6D_{1122}x_1^2x_2^2 \nn
    &&+4D_{1222}x_1x_2^3 + D_{2222}x_2^4,
\label{s:ddix}
\ea
and its partial contributions have to be evaluated numerically.
The terms $D_{1111}$ and $D_{2222}$ can be calculated
through the expression of the fourth virial coefficient for a
monodisperse fluid of particles with diameter $\sigma_{1}$ or
$\sigma_{2}$, respectively, \emph{i.e.},
\be
D_{1111}=\frac{\pi^3}{64}b_4\sigma_1^6,
\ee
\be
D_{2222}=\frac{\pi^3}{64}b_4\sigma_2^6,
\ee
where $b_4=8 (2 + 10/\pi^2 - 9 \sqrt{3}/2 \pi)\simeq 4.25785$. On the other hand, the coefficients $D_{1112}$ and
$D_{1122}$ are cluster integrals which are represented by the
following four-point color graphs:
\ba
D_{1112}&=&-\frac{1}{8}\left(3\DA+3\DB+3\DC\right.\nn
&&\left.+\DD\right),
\label{s:d1112}
\ea
\ba
D_{1122}&=&-\frac{1}{8}\left(2\DE+\DF+4\DG\right.\nn
&&\left.+\DI+\DK+\DL\right).
\label{s:d1122}
\ea
The open and solid circles in each graph identify
particles belonging to species 1 and 2, respectively. Each bond
contributes a factor to the integrand in the form of a Mayer step
function. Space integration is carried out over all the vertices
of the graph. Of course, the coefficient $D_{1222}$ is obtained from Eq.\ \eqref{s:d1112} by exchanging the open and solid circles.

For later use, let $g_{ij}(\rho)$ be
the values of the radial distribution functions at contact of the NAHD mixture. This quantity is related to the pressure via the virial equation of state\cite{HM06}
\be
\beta P=\rho+\frac{\pi}{2}\rho^2\sum_{i,j=1}^2 x_i x_j \sigma_{ij}^2 g_{ij}(\rho).
\label{virial}
\ee
No
general expression is known for $g_{ij}(\rho)$, but it may formally
be expanded in a power series in density as
\be
g_{ij}(\rho)=1+\frac{\pi}{4}\rho \sum_{k=1}^{2}x_{k}c_{k;ij}+\frac{\pi^2}{16}\rho^2\sum_{k,\ell=1}^{2}x_{k}x_\ell d_{k\ell;ij}
 +\cdots,
\label{1M}
\ee
where the coefficients $c_{k;ij}$, $d_{k\ell;ij}$, \ldots are
independent of the mole fractions but in general depend in a non-trivial way on the set of diameters $\{\sigma_{ij}\}$. Only the coefficients linear in $\rho$ (\emph{i.e.},
$c_{k;ij}$) are known analytically ({\it cf.} Refs.\ \ocite{AEH99,SHY05}), namely
\be
c_{1;11}=\frac{b_3}{2}\sigma_{1}^2,
\label{H.2}
\ee
\be
c_{2;11}=\frac{b_3}{2}\sigma_{1}^2
G\left(\frac{\sigma_{12}}{\sigma_{1}}\right),\quad
c_{1;12}=\frac{b_3}{2}\sigma_{1}^2
H\left(\frac{\sigma_{12}}{\sigma_{1}}\right).
\label{H.2b}
\ee
Other combinations of indices follow from the exchange of indices $1$ and $2$ in the above results. We recall that the functions $G(x)$ and $H(x)$ are given by Eqs.\ \eqref{H.3}--\eqref{H.4b}. In fact, insertion of Eq.\ \eqref{1M} into Eq.\ \eqref{virial} yields
\begin{equation}
C_{ijk}=\frac{\pi^2}{24}\left(c_{k;ij}\sigma_{ij}^2+
c_{j;ik}\sigma_{ik}^2+c_{i;jk}\sigma_{jk}^2\right),
\label{n2}
\end{equation}
so that Eqs.\ \eqref{H.9a}--\eqref{H.9d} are recovered from Eqs.\ \eqref{H.2} and \eqref{H.2b}

\section{Approximate theoretical approaches}
\label{sec3}
Before we evaluate numerically the partial fourth virial coefficients, let us recall the approximate results derived for them with different theoretical approaches. These were presented in a unified framework within the description of general multi-component nonadditive hard-sphere mixtures in $d$ dimensions. We will consider here the particular case of a binary mixture in two dimensions, only quote the relevant results, and refer the interested reader to Ref.\ \ocite{SHY10} for details.

\subsection{MIX1 approximation}

In the so-called MIX1 theory for NAHD mixtures, which we will label with a superscript M, the fourth virial coefficients are given by
\ba
D_{ijk\ell}^{\text{M}}&=&\frac{\pi^3}{192}\left[\left(\frac{\sigma_{i}+\sigma_{j}}{2}\right)^2d^{\text{add}}_{k\ell;ij}\left(1+3Y_{ij}^\text{M}\right)\right.\nn
&& +\left(\frac{\sigma_{i}+\sigma_{k}}{2}\right)^2d^{\text{add}}_{j\ell;ik}\left(1+3Y_{ik}^\text{M}\right)\nn
&& +\left(\frac{\sigma_{i}+\sigma_{\ell}}{2}\right)^2d^{\text{add}}_{jk;i\ell}\left(1+3Y_{i\ell}^\text{M}\right)\nn
&& +\left(\frac{\sigma_{j}+\sigma_{k}}{2}\right)^2d^{\text{add}}_{i\ell;jk}\left(1+3Y_{jk}^\text{M}\right)\nn
&& +\left(\frac{\sigma_{j}+\sigma_{\ell}}{2}\right)^2d^{\text{add}}_{ik;j\ell}\left(1+3Y_{j\ell}^\text{M}\right)\nn
&&\left. + \left(\frac{\sigma_{k}+\sigma_{\ell}}{2}\right)^2d^{\text{add}}_{ij;k\ell}\left(1+3Y_{k\ell}^\text{M}\right)\right]
.
\label{1.7}
\ea
In Eq.\ (\ref{1.7}), $d^{\text{add}}_{k\ell;ij}$ are the second-order coefficients defined in Eq.\ \eqref{1M}, particularized to the additive case ($\Delta=0$). Here we adopt the approximation\cite{SYH99,SYH01,HYS02}
\be
d^{\text{add}}_{k\ell;ij}=\sigma_k^2\sigma_\ell^2\left[1+\left(\frac{b_4}{2}-1\right)\frac{\sigma_i\sigma_j}{\sigma_i+\sigma_j}
\frac{\sigma_k+\sigma_\ell}{\sigma_k\sigma_\ell}\right].
\ee
Moreover, in Eq.\ \eqref{1.7},
\be
Y_{ij}^\text{M}\equiv 2\Delta(1-\delta_{ij}),
\label{1.2}
\ee
where $\delta_{ij}$ is the Kronecker delta.

\subsection{Paricaud's modified MIX1 theory (mMIX1)}

In the generalization of Paricaud's approximation that was made in Ref.\ \ocite{SHY10}, which will be identified with the superscript mM, and restricting the result to two-dimensional binary mixtures, the partial composition-independent fourth virial coefficients have the same form as in the MIX1 approximation but one has to replace $Y_{ij}^\text{M}$ with $Y_{ij}^\text{mM}$, where this latter is given by
\be
Y_{ij}^\text{mM}\equiv \Delta(2+\Delta)(1-\delta_{ij}).
\label{1.8}
\ee

\subsection{Hamad's proposal}

In the work of Hamad and his collaborators,\cite{AEH99,HY00} denoted here by the superscript H, the fourth virial coefficients are given by
\ba
{D}_{ijk\ell}^{\text{H}}&=&\frac{\pi^3b_4}{96 b_3^2}\left(\sigma_{ij}^2
c_{k;ij}c_{\ell;ij}+\sigma_{ik}^2
c_{j;ik}c_{\ell;ik}+\sigma_{i\ell}^2
c_{j;i\ell}c_{k;i\ell}\right.\nn &&\left.+\sigma_{jk}^2
c_{i;jk}c_{\ell;jk}+\sigma_{j\ell}^2
c_{i;j\ell}c_{k;j\ell}+\sigma_{k\ell}^2
c_{i;k\ell}c_{j;k\ell}\right).\nn
\label{2.6}
\ea

\subsection{The Santos-L\'opez de Haro-Yuste proposal}

In the proposal made in 2005 by three of us,\cite{SHY05} hereafter denoted by the superscript SHY,  the fourth virial coefficients are expressed in terms of the partial second and third composition-independent virial coefficients and of $b_3$ and $b_4$. Written for $d=2$ they read
\begin{eqnarray}
D_{ijk\ell}^{\text{SHY}}&=&\frac{\pi(b_4-2)}{16(b_3-2)}\left(\sigma_{i}^2
C_{jk\ell}+\sigma_{j}^2 C_{ik\ell}+\sigma_{k}^2
C_{ij\ell}+\sigma_{\ell}^2 C_{ijk}\right)\nn
&&
-\frac{\pi^2(b_4-b_3)}{96(b_3-2)}\left(\sigma_i^2 \sigma_j^2
B_{k\ell}+\sigma_{i}^2 \sigma_{k}^2 B_{j\ell}+\sigma_{i}^2
\sigma_{\ell}^2 B_{jk}\right.\nonumber\\
&&\left.+\sigma_j^2 \sigma_k^2 B_{i\ell}+\sigma_{j}^2 \sigma_{\ell}^2
B_{ik}+\sigma_{k}^2 \sigma_{\ell}^2 B_{ij}\right).
\label{5.2}
\end{eqnarray}

\section{Results}
\label{sec4}

\begin{figure}
\centering
\includegraphics[width=\widthfigure\columnwidth]{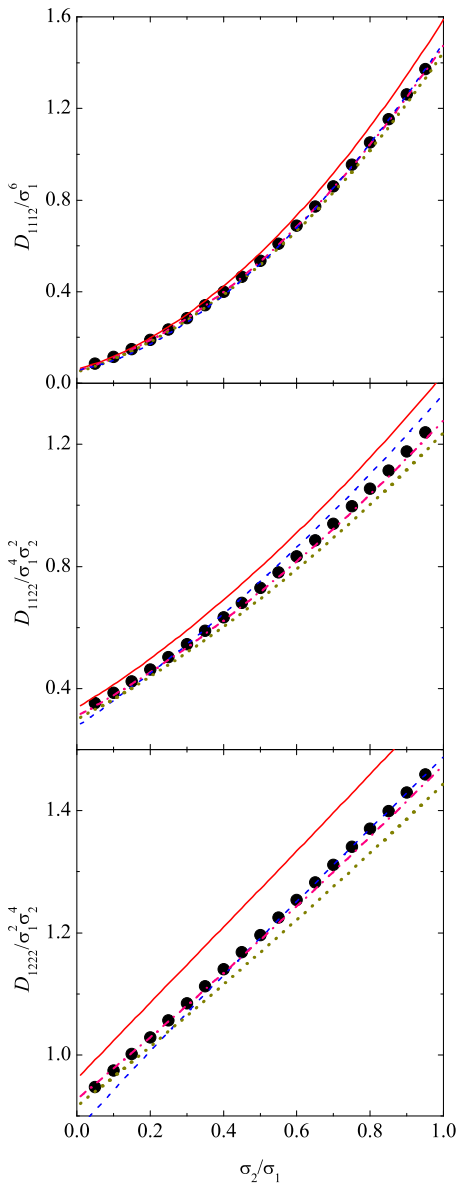}
\caption{Plot of the composition-independent fourth virial
coefficients $D_{1112}$, $D_{1122}$ and $D_{1222}$ versus the size
ratio $q=\sigma_{2}/\sigma_{1}$ for a nonadditivity parameter
$\Delta=-0.1$. The dotted {(green)} lines correspond to the original
MIX1 theory, Eq.\ \protect\eqref{1.7}, the  dash-dot {(pink)} lines correspond to the mMIX1
theory, Eq.\ \protect\eqref{1.7}, with $Y_{ij}^\text{M}\to Y_{ij}^\text{mM}$,  the dashed {(blue)} lines correspond to
Hamad's proposal,  Eq.\ \protect\eqref{2.6}, and the solid {(red)} lines correspond to the SHY  proposal, Eq.\ \protect\eqref{5.2}. The symbols are our MC data.\label{fig1}}
\end{figure}
\begin{figure}
\centering
\includegraphics[width=\widthfigure\columnwidth]{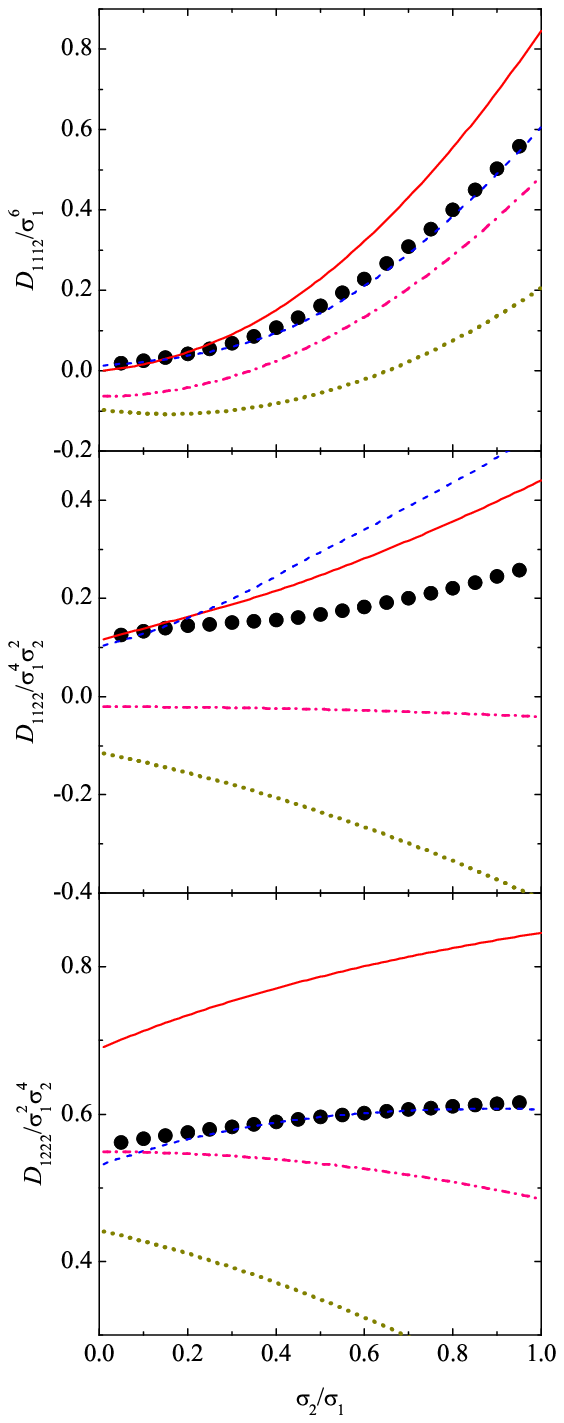}
\caption{Same as in  Fig.\ \protect\ref{fig1}, but for $\Delta=-0.3$.
\label{fig2}}
\end{figure}
\begin{figure}
\centering
\includegraphics[width=\widthfigure\columnwidth]{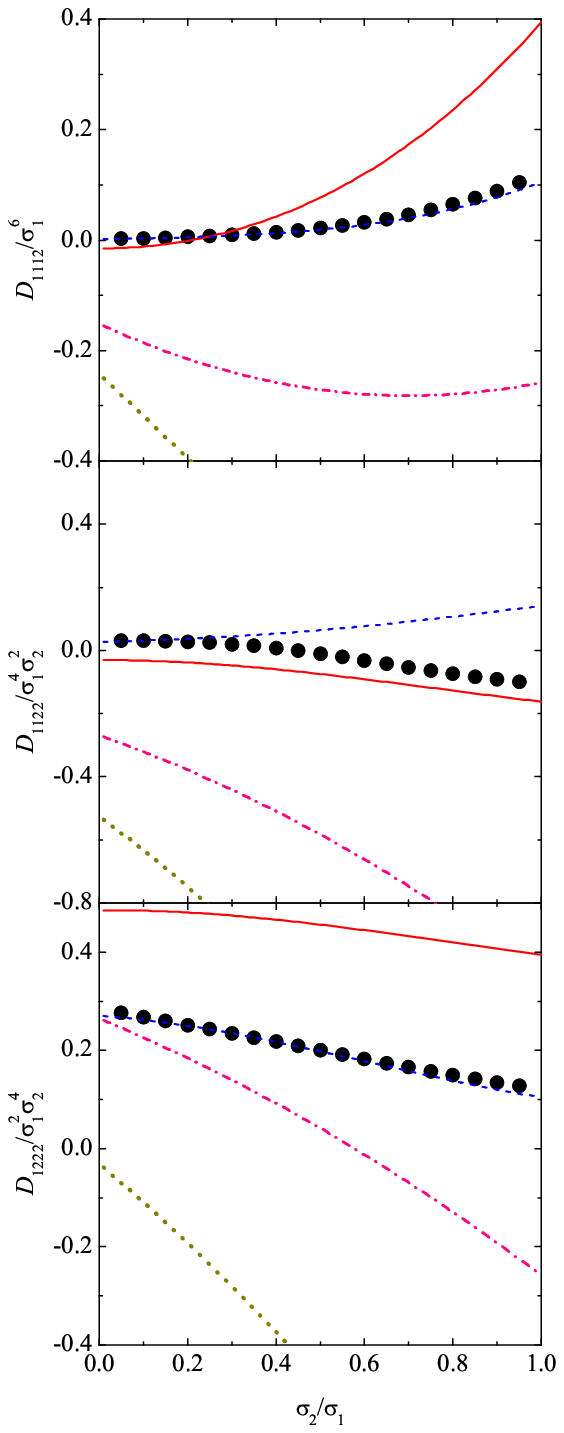}
\caption{Same as in  Fig.\ \protect\ref{fig1}, but for $\Delta=-0.5$.
\label{fig3}}
\end{figure}
\begin{figure}
\centering
\includegraphics[width=\widthfigure\columnwidth]{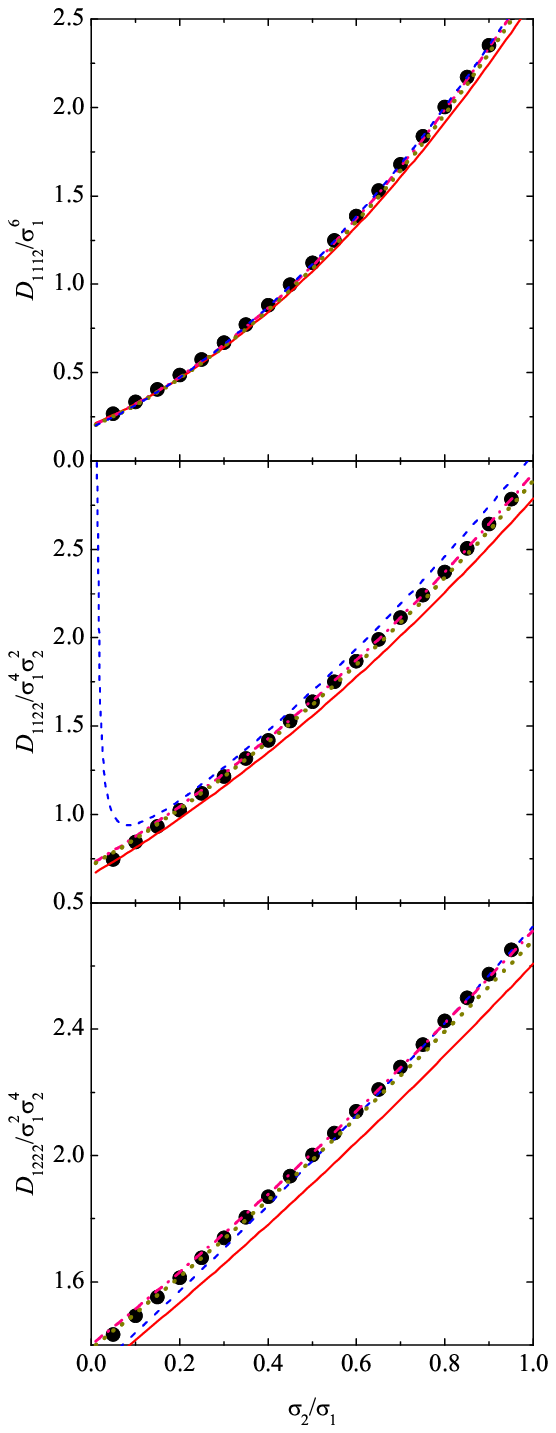}
\caption{Same as in  Fig.\ \protect\ref{fig1}, but for $\Delta=0.1$.
\label{fig4}}
\end{figure}
\begin{figure}
\centering
\includegraphics[width=\widthfigure\columnwidth]{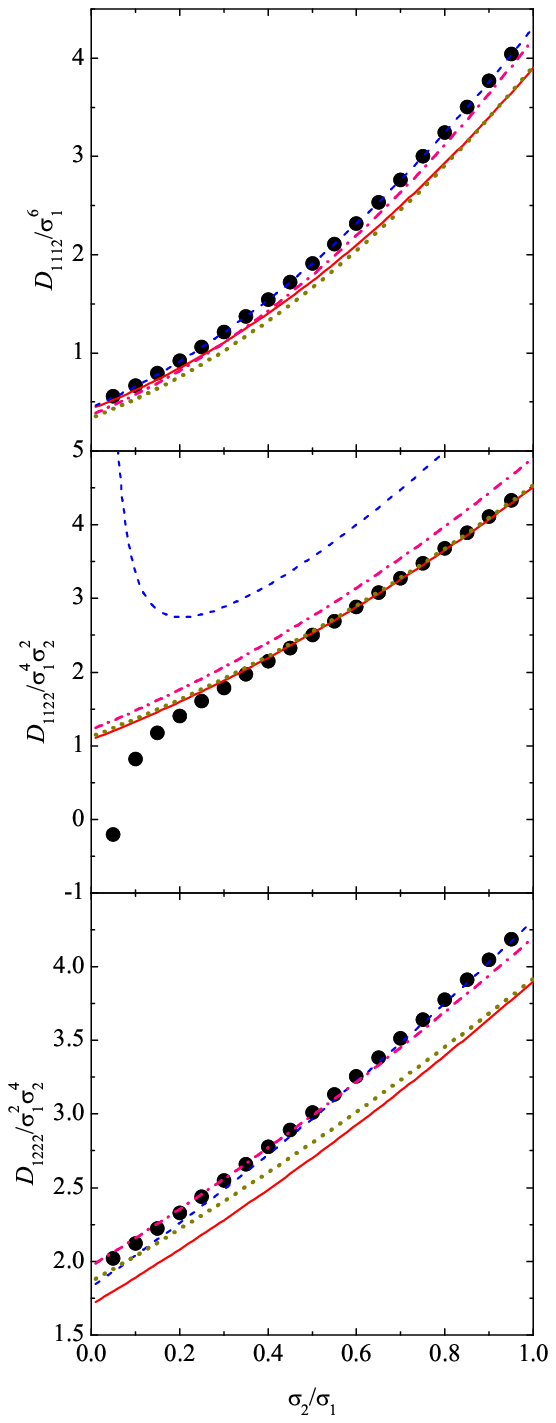}
\caption{Same as in Fig.\ \protect\ref{fig1}, but for $\Delta=0.3$.
\label{fig5}}
\end{figure}
\begin{figure}
\centering
\includegraphics[width=\widthfigure\columnwidth]{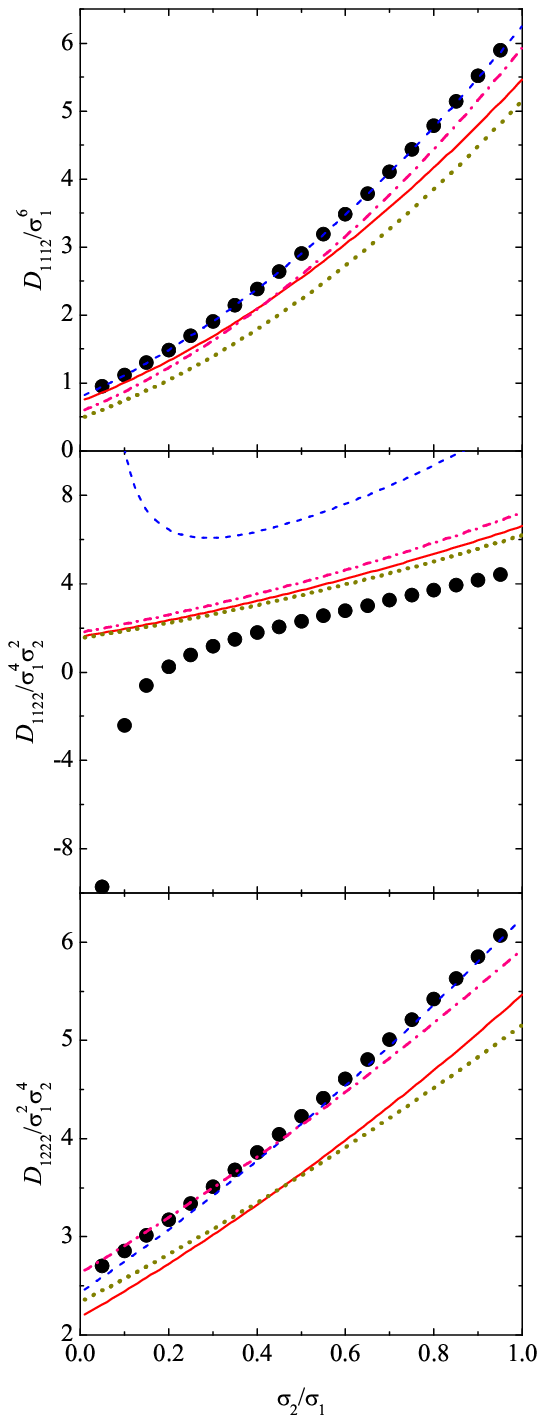}
\caption{Same as in Fig.\ \protect\ref{fig1}, but for $\Delta=0.5$.
\label{fig6}}
\end{figure}

In this section we report the results of our calculations. In order to evaluate the irreducible cluster integrals which enter the expression of the composition-independent coefficients
$D_{ijk\ell}$ [see Eqs.\ \eqref{s:d1112} and \eqref{s:d1122}],
we used a standard MC integration procedure. The algorithm
produces a significant set of configurations which are compatible with
the Mayer graph one wants to evaluate. We first fix particle 1 of species
$i$ at the origin and sequentially deposit the
remaining three particles at random but in such a way that
particle $\alpha+1$ overlaps with particle $\alpha$ (where $\alpha=0, 1,2,3$).
This procedure generates an open chain of overlapping particles which
is taken as a ``trial configuration''. A ``successful configuration'' is a
closed-chain configuration \emph{(i.e.}, a configuration in which particle $1$
further overlaps with particle $4$) where, moreover, the residual cross-linked
``bonds'' which are present in the Mayer graph that is being calculated
are also retrieved. The ratio of the number of successful configurations
($N_s$) to the total number of trial configurations ($N_t$)
yields asymptotically the value of the cluster integral
relative to that of the open-chain graph
which, in turn, is trivially related to a product of the partial second-order
virial coefficients $B_{ij}$.\cite{BN86,B92}
The numerical accuracy of the MC results
obviously depends on the total number of trial configurations. The error on
the cluster integral $J$ is estimated as:\cite{K77}
\begin{equation}
\text{error} = \left[ \frac {J(J-1)}{N_t} \right]^{1/2}.
\label{s:error}
\end{equation}
However, as a result of the accumulation of statistically independent
errors, the global uncertainty affecting the partial virial coefficients
is higher than the error estimated for each cluster integral that enters
the expression of $D_{ijk\ell}$.
A typical MC run consisted of $4\times10^9$ independent moves.
The error on each cluster integral, as estimated
through Eq.\ \eqref{s:error}, turned out to be systematically
less than $0.05\%$, with a cumulative uncertainty on the
partial virial coefficients lower than $0.5\%$.

{The numerical values of $D_{1112}/\sigma_1^6$, $D_{1122}/\sigma_1^6$, and $D_{1222}/\sigma_1^6$  for $\Delta=\pm 0.05$, $\pm 0.1$, $\pm 0.2$, $\pm 0.3$, $\pm 0,4$, and $\pm 0.5$ and $q=0.05,0.10,\ldots,0.90,0.95$ are presented in tabular form in the supplementary material to this paper.\cite{note_12_04}}

Now we proceed to assess the merits of the different theoretical formulae for the composition-independent partial fourth virial coefficients that we presented in section\ \ref{sec3}. For that purpose, although we have made an exhaustive analysis, in Figs.\ \ref{fig1}--\ref{fig6} we present only some illustrative cases in which we compare the performance of the different approximations against the MC data. The graphs corresponding to the other values of $\Delta$ that appear {in the tables of the supplementary material to this paper\cite{note_12_04} are available upon request.}

From these figures it is clear that, overall, the proposal by Hamad,\cite{AEH99} Eq.\ (\ref{2.6}), is very good for $D_{1112}$ and $D_{1222}$ but rather bad for $D_{1122}$ if $|\Delta|>0.1$, irrespective of the value (positive or negative) of $\Delta$. None of the theories shows a good performance in the case of $D_{1122}$  but at least the SHY proposal leads to reasonable quantitative agreement in the positive region of this coefficient, being particularly superior to all other approximations for negative values of $\Delta$.

\section{Equation of state}
\label{EOS}

\begin{figure}
\centering
\includegraphics[width=\widthfigure\columnwidth]{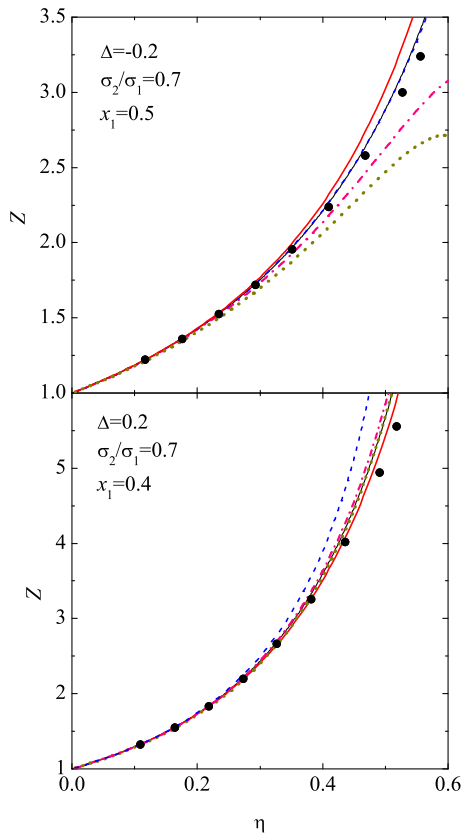}
\caption{{Plot of the compressibility factor $Z$ versus the total packing fraction $\eta$ for {NAHD} mixtures with $\Delta=-0.2$, $\sigma_2/\sigma_1=0.7$, $x_1=0.5$ (top panel) and $\Delta=0.2$, $\sigma_2/\sigma_1=0.7$, $x_1=0.4$ (bottom panel). The dotted {(green)} lines correspond to the original
MIX1 theory,  the  dash-dot {(pink)} lines correspond to the mMIX1
theory,   the dashed {(blue)} lines correspond to
Hamad's proposal,   the thick solid {(red)} lines correspond to the SHY  proposal,  and the thin solid {(black)} lines correspond to the {RVE, Eq.\ \protect\eqref{rve}}. The symbols are our MC data.\protect\cite{note_12_04_2} Note that the RVE and the Hamad curves are practically indistinguishable in the top panel, while the RVE and the MIX1 curves are practically indistinguishable in the bottom panel.}\label{fig7}}
\end{figure}

{
Since the convergence of the virial expansion is unknown and truncating the series after the first four terms would not guarantee a satisfactory outcome, in this section we will use the knowledge of the first four virial coefficients to illustrate the performance of a well established approach to the equation of state of  fluids that incorporates such knowledge. Hence we will consider the rescaled virial expansion (RVE) proposed by Baus and Colot\cite{BC87,BXHB88} to obtain an (approximate) equation of state for an asymmetric {NAHD} mixture.
The RVE equation of state truncated to the fourth order has the following form:
\begin{equation}
{Z\equiv \frac{\beta P}{\rho}=\frac{1 + c_1\eta + c_2\eta^{2} + c_3\eta^{3}}{(1 - \eta)^{2}},}
\label{rve}
\end{equation}
{where $Z$ is the compressibility factor, $\eta=\rho\xi$, with $\xi\equiv (\pi/4)(x_1\sigma_1^2+x_2\sigma_2^2)$, is the total packing fraction, and the coefficients $c_1$, $c_2$, and $c_3$} are obtained by identification
with the corresponding coefficients which show up in the virial series. Specifically,
in the present case one has
\begin{equation}
{c_{1} = \frac{B}{\xi} - 2,\quad c_2=\frac{C}{\xi^2}-2\frac{B}{\xi}+1,\quad c_3=\frac{D}{\xi^3}-2\frac{C}{\xi^2}+\frac{B}{\xi}.}
\end{equation}

In Fig.\ \ref{fig7} we present an illustrative comparison between the results for the compressibility factor of {two}  binary {NAHD} mixtures as a function of the packing fraction as derived from the RVE, {Eq.\ \eqref{rve},} and those obtained by MC simulation.\cite{note_12_04_2} In {both mixtures} the size ratio is $q=0.7$ and  a negative nonadditivity $\Delta=-0.2$ {(with $x_1=0.5$)} and a positive nonadditivity $\Delta=0.2$ {(with $x_1=0.4$)} have been considered. For comparison, the results stemming out of the compressibility factors  corresponding to the different theoretical approximations mentioned in section \ref{sec3}  are also included in this figure.
As discussed in Ref.\ \protect\ocite{SHY10}, for the actual calculations using the compressibility factors corresponding to the different theoretical approaches, one needs to specify the contact values of the one-component system for the Hamad and the SHY approaches and those of an \emph{additive} hard-disk mixture in the MIX1 and mMIX theories. For the former we have used an accurate proposal by {Luding,\cite{L01b,LS04}  while for the latter we have considered the quadratic approximation proposed in Ref.\ \ocite{SYH02}, complemented with Luding's one-component value.\cite{LS04}}

It is clear that in the case of {the mixture with} negative nonadditivity, the best agreement is provided by both the RVE and the Hamad compressibility factor, followed by the SHY compressibility factor. In fact, the former two are hardly distinguishable. On the other hand, for positive nonadditivity it is the SHY compressibility factor the one that provides the best agreement, followed by  both the RVE and the MIX1 compressibility factor. These latter two are virtually indistinguishable.
}

\section{Concluding remarks}
\label{sec5}

In this paper we have reported MC calculations of the fourth virial coefficients of asymmetric NAHD mixtures over a rather wide range of size ratios $q$ and values of the nonadditivity parameter $\Delta$. These results complement those reported earlier\cite{S11} for symmetric mixtures {($q=1$)} and, {as illustrated in the case of the RVE and the {mixtures} discussed in  Sec.\ \ref{EOS},} may prove useful for the development of new equations of state for NAHD mixtures. {In particular, one could also consider using the availability of the fourth virial coefficients provided in this paper to derive another approximation to the compressibility factor of asymmetric {NAHD} mixtures via the $y$-expansion} proposed by Barboy and Gelbart.\cite{BG79,BG80} Here we have {mainly} used the data to assess the merits of different theoretical approaches leading to the thermodynamic properties of NAHD mixtures {with respect to their performance in the prediction of the values of the fourth virial coefficients}.

One immediate conclusion is that none of the existing theories can account for all the features observed in the MC data. In contrast with what happened in NAHS mixtures,\cite{SHY10} here the theoretical approach by Hamad\cite{AEH99} outperforms all the rest. In this regard, it is somewhat striking that its very good performance concerning $D_{1112}$ and $D_{1222}$ is not also found for $D_{1122}$, where the SHY proposal does the best overall job.  In any case, the comparison we have presented is only indicative of the performance with respect to the fourth virial coefficients, but the full assessment will have to do with the compressibility factor and with the issue of fluid-fluid demixing. We plan to address these points in the near future.

\begin{acknowledgments}
Two of us (A.S. and S.B.Y) acknowledge the financial support of the Spanish government  (Grant No.\ FIS2010-16587) and  the Junta de Extremadura (Spain) (Grant No.\ GR10158) (partially financed by FEDER funds). The work of M.L.H. has been partially supported by DGAPA-UNAM under project IN-107010-2.
\end{acknowledgments}

\bibliographystyle{apsrev}
\bibliography{D:/Dropbox/Public/bib_files/liquid}

\end{document}